\documentclass[aps,prb,twocolumn,superscriptaddress]{revtex4-1}
\usepackage[dvips]{epsfig}

\usepackage[sort&compress]{natbib}

\begin{document}

%\preprint{}

\title{Sub-nanosecond Electro-optic Modulation of Triggered Single Photons from a Quantum Dot}


\author{Matthew T. Rakher} \email{matthew.rakher@gmail.com}
\altaffiliation{Current Address: Departement Physik, Universit\"{a}t Basel, Klingelbergstrasse 82, CH-4056 Basel, Switzerland}
\author{Kartik Srinivasan}
\affiliation{Center for Nanoscale Science and Technology, National
Institute of Standards and Technology, Gaithersburg, MD 20899, USA}
%\thanks{}


\date{\today}

\begin{abstract}
Control of single photon wave-packets is an important resource for developing hybrid quantum systems which are composed of different physical systems interacting via photons.  Here we extend this control to triggered photons emitted by a quantum dot, temporally shaping single photon wave-packets on timescales fast compared to their radiative decay by electro-optic modulation.  In particular, telecommunications-band single photons resulting from the recombination of an exciton in a quantum dot with exponentially decaying wave-packets are synchronously modulated to create Gaussian-shaped single photon wave-packets.  We explore other pulse-shapes and investigate the feasibility of this technique for increasing the indistinguishability of quantum dot generated single photons.
\end{abstract}

%\keywords{}

\maketitle

Single photons are an integral part of many protocols in quantum information science such as quantum key distribution \cite{ref:Gisin_RMP_02} and quantum computation \cite{ref:Knill,ref:Raussendorf_PRL01}.  One of the most promising sources of single photons are single, self-assembled quantum dots (QDs)\cite{ref:Shields_NPhot}.  Because they can be formed in commonly-used optoelectronic materials, they also offer the ability to control their emission properties by designing monolithic cavity \cite{ref:Santori,ref:Strauf_NPhot} or waveguide micro-structures \cite{ref:Claudon,ref:Davanco2}.  The temporal shape of the wave-packet of these photons is determined by the recombination process of carriers confined in the QDs which usually results in an exponentially-decaying amplitude over a timescale of 0.5~ns to 2~ns.  However, this pulse shape is not ideal for interacting with other two-level quantum systems such as atoms, ions, or other QDs \cite{ref:Cirac,ref:Gorshkov_PRL07} as part of a large-scale quantum network \cite{ref:Kimble_Nat08}.  In this Letter, we perform amplitude modulation of single photons from a single quantum dot in the telecommunications-band to create Gaussian-shaped wave-packets and other non-trivial pulse shapes by synchronized, sub-nanosecond, electro-optic modulation.  Previously, single photon electro-optic amplitude and phase modulation have been performed on single photons emitted by an atomic ensemble\cite{ref:Kolchin_PRL_08} and by a trapped atom in a cavity\cite{ref:Specht_NPhot_09}, but in both cases the photon wave-packet was $\approx$150~ns in width.  In addition, control of single photon waveforms has been demonstrated using a lambda-type level system in an atom\cite{ref:McKeever} and an ion\cite{ref:Keller}, but this technique would be difficult to apply to a QD.  We extend electro-optic modulation to QD-generated single photons and reduce the required modulation timescales by more than two orders of magnitude, leading to a robust and flexible single photon source capable of efficient interactions in a quantum network.

\begin{figure}
\centering
        \includegraphics[width=8.5cm, clip=true]{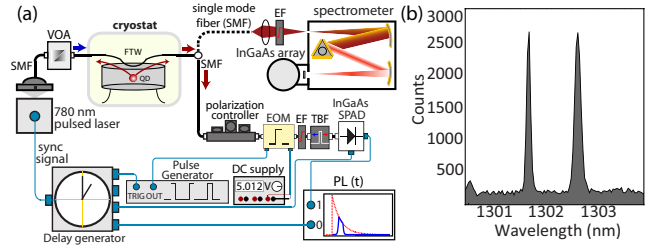}
    \caption{(a) Schematic of the experimental setup for the generation and manipulation of triggered signal photons from a quantum dot.  Definition of acronyms: VOA=variable optical attenuator, FTW=fiber taper waveguide, EOM=electro-optic modulator, TBF=tunable bandpass filter, SPAD=single photon counting avalanche photodiode, EF=edgepass filter, PL = photoluminescence. (b) PL spectrum of the emission of the single QD used in this work.  The spectrum was taken for an integration time of 60~s under an excitation power of 10~nW.  }
    \label{fig:fig1}
\end{figure}

Single photons at 1.3~$\mu$m are generated by photoluminescence (PL) from a single InAs quantum dot.  The QDs used in this work are grown by molecular beam epitaxy where they are made to self-assemble in an InGaAs quantum well contained in a 256~nm thick GaAs device layer.  The sample is etched to create isolated mesas of $\approx$2.5~$\mu$m diameter in the device layer by a nanofabrication process consisting of electron-beam lithography, optical lithography, and wet and dry etching.  This isolation enables efficient, near-field optical coupling to a fiber taper waveguide (FTW) \cite{ref:Srinivasan15}.  For this work, the FTW is used for efficient excitation and collection of emission from QDs into single mode fiber, which has been estimated to be $\approx$0.1~$\%$ in previous experiments\cite{ref:Srinivasan15,ref:Rakher_NPhot_2010}.  As shown in Fig.~\ref{fig:fig1}(a), the sample containing QDs resides in a liquid-Helium flow cryostat at a temperature of $\approx$7~K.  Optical excitation by a 50~ps, 780~nm pulsed laser diode operating at 50~MHz is introduced into the cryostat by fiber-optic feedthroughs after attenuation by a variable optical attenuator (typical excitation powers were $\approx$10~nW).  Subsequent carrier relaxation and recombination inside the QDs generates photoluminescence (PL) which is efficiently captured by the FTW.  As depicted in Fig.~\ref{fig:fig1}(a), the fiber output of the cryostat can then be connected to a monochromator with a liquid N$_2$-cooled InGaAs array after spectral filtering by an edge pass filter to remove excess excitation light.  This enables PL spectroscopy to identify a candidate QD that is both spectrally and spatially separated from nearby QDs.  Figure~\ref{fig:fig1}(b) depicts such a PL spectrum of the QD used in this work and shows two sharp transitions, corresponding to the positively charged exciton, $X^+$, near 1301.5 nm and the neutral exciton, $X^0$, near 1302.5 nm.  This identification is based on polarization-resolved spectroscopy and is not shown here.  Furthermore, in Ref.~\onlinecite{ref:Rakher_NPhot_2010} this QD was explicitly shown through photon antibunching experiments to emit single photons and the measurement is not repeated here.

\begin{figure}
\centering
        \includegraphics[width=8.5cm, clip=true]{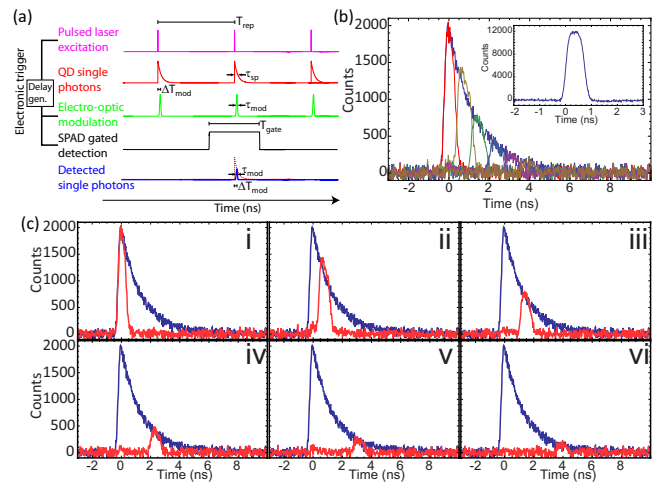}
    \caption{(a) Schematic of the timing sequence of the experiment.  $T_{rep}$ corresponds to the repetition period of the excitation laser, $\tau_{sp}$ is the spontaneous emission lifetime, $\tau_{mod}$ is the width of the electro-optic modulation, $\Delta T_{mod}$ is the delay of the modulation with respect to the incoming photon, and $T_{gate}$ is the width of the detection window.  (b) Unmodulated and modulated single photon waveforms for different modulation delays. The temporal profile of the modulation is shown in the inset and corresponds to a full-width half-maximum of 720~ps~$\pm$~18~ps. (c) Same traces shown in (b) but separated for clarity.  The modulation delays are $\Delta T_{mod}=\{$0.0, 0.8, 1.6, 2.4, 3.2, 4.0$\}$~ns respectively.  The unmodulated single photon waveform is shown for comparison in blue.  All traces are integrated for 1200~s. }
    \label{fig:fig2}
\end{figure}

Electro-optic modulation is performed by directing the single photon PL into a fiber-coupled, LiNbO$_3$ electro-optic intensity modulator (EOM) after polarization manipulation.  The DC bias of the EOM was controlled by a low-noise DC supply while the rf input was connected to an externally-triggered pulse generator capable of producing pulses as short as $\tau_{mod}\approx$300~ps with sufficient amplitude to reach the $V_{\pi}\approx4$~V of the EOM.  Of critical importance is the synchronization of the EOM to the
incoming single photon pulses.  While in Ref.
\onlinecite{ref:Kolchin_PRL_08} this was done through detection of a
heralded photon of a biphoton pair, here the repetition period, $T_{rep}$, of
the 780~nm pulsed laser excitation source determines the single
photon arrival times as shown schematically in Fig.~\ref{fig:fig2}(a).  Thus, the timing of the EOM pulse is set to
overlap with the arrival of the incoming photon at a specified time by using a delay generator
triggered by the electronic synchronization signal of the excitation laser.  The delay generator serves as a master clock (Fig.~\ref{fig:fig1}), allowing the electro-optic modulation to be controllably delayed with respect to the incoming photon by an amount $\Delta T_{mod}$.  It also ensures synchronization with respect to the 50~ns detection window of the telecommunications-band InGaAs/InP single photon counting avalanche photodiode (SPAD), which must be operated in a gated detection mode to avoid strong afterpulsing \cite{ref:Hadfield_nphoton_09} in contrast to Si-based SPADs used for shorter wavelength detection.  The SPAD gate is triggered at 5 MHz by a signal from the delay generator so as to coincide with the arrival of every tenth optical pulse, which at this point has been modulated and spectrally isolated using a tunable bandpass filter so that only the single photon emission from the $X^0$ transition at 1302.5 nm is present.  Using an electronic pulse from the
delay generator and the electronic pulse from the SPAD ($\approx250$~ps timing jitter), a time
resolved histogram of detection events can be formed using a
time-correlated single photon counting system.

Before directing the QD single photons into the modulation setup, we first synchronize the electronics and characterize the temporal width of the modulation by using a continuous-wave laser at 1302.5~nm, the same wavelength as the neutral exciton transition.  This laser was attenuated and passed through the EOM and
measured by the SPAD.  The temporal
histogram of the resulting waveform is shown in the inset of Fig.~\ref{fig:fig2}(b)
as measured by the time correlation system.  The extracted
full-width-half-maximum (with uncertainty given by the 95~$\%$ confidence level from fit) of the Gaussian-like pulse for this rf setting was 720~ps~$\pm$~18~ps with an extinction ratio of $\gtrsim20$~dB.  Next, the temporal shape of the QD single photon
wave-packet was measured under no modulation.  In this case, the
pulse generator was turned off and the DC bias was set to maximum
transmission.  The trace was integrated for 1200~s and is shown in blue in Fig.~\ref{fig:fig2}(b) and Fig.~\ref{fig:fig2}(c).  The curve is a single
exponential decay with time constant 1.4~ns~$\pm$~0.1~ns.

To modulate the exponentially-decaying single photon wave-packet, the EOM was setup exactly as in the inset of Fig.~\ref{fig:fig2}(b).  The delay between the arrival
of the QD single photon and the triggering of the EOM, $\Delta T_{mod}$, was varied in
intervals of 0.8 ns and the resulting waveforms were integrated for 1200~s.  The histograms are shown together in Fig.~\ref{fig:fig2}(b) and separated in Fig.~\ref{fig:fig2}(c) in red, corresponding to temporal delays
of $\{$0.0, 0.8, 1.6, 2.4, 3.2, 4.0$\}$~ns respectively.  The
modulated waveform heights nicely follow the contour of the
exponential decay shown in blue. This data clearly demonstrates the
modulation of triggered single photon wave-packets from an exponentially-decaying amplitude to a more convenient Gaussian-shaped pulse.

\begin{figure}
\centering
        \includegraphics[width=8.5cm, clip=true]{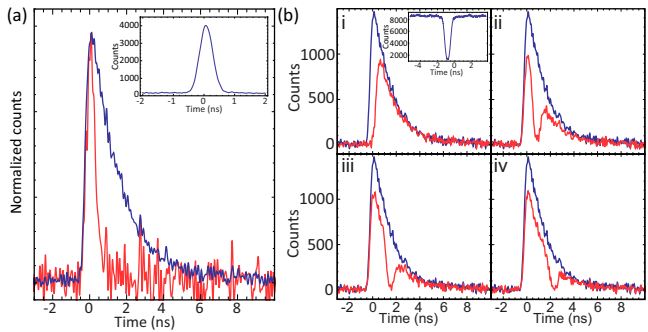}
    \caption{(a) Single photon modulated (red) waveform for a modulation width, $\tau_{mod}$, of 520~ps~$\pm$~13~ps along with the unmodulated waveform (blue) for comparison.  The temporal profile of the modulation is shown in the inset. (b) Single photon modulated waveforms (red) for the inverted Gaussian modulation profile ($\tau_{mod}$=770~ps~$\pm$~19~ps) shown in the inset of (i) for modulation delays of $\Delta T_{mod}=\{$0.0, 0.8, 1.6, 2.4$\}$~ns respectively.  The unmodulated waveform (blue) is shown for comparison.}
    \label{fig:fig3}
\end{figure}

The flexibility of our setup also easily allowed for other waveforms by simply changing the settings of the pulse generator.  Figure~\ref{fig:fig3}(a) shows the modulated (red) and unmodulated (blue) single-photon waveforms for the amplitude modulation shown in the inset.  In this case, the modulation profile was a Gaussian with width $\tau_{mod}$=520~ps~$\pm$~13~ps and each trace was integrated for 1200~s.  The modulation profile could also be inverted, as shown in the inset of the first panel of Fig.~\ref{fig:fig3}(b), to create a Gaussian-shaped notch of width $\tau_{mod}=$770~ps~$\pm$~19~ps.  The resulting single photon waveforms under this modulation are shown in red in the panels of Fig.~\ref{fig:fig3}(b) for modulation delays of $\Delta T_{mod}=\{$0.0, 0.8, 1.6, 2.4$\}$~ns respectively.  More complex waveforms are possible provided the pulse generator and electro-optic modulator are capable of producing them.  In this work, the timing was limited by the pulse generator to $\approx$300~ps but the 10 GHz EOM could in principle reach pulse widths on the order of 100~ps and 40 GHz EOMs are readily available at 1300~nm.  Such waveforms could be useful to encode more information in a single photon \cite{ref:Broadbent_PRA09} or to hide and recover a single photon in a noisy signal \cite{ref:Belthangady_PRL10}.

It is important to note that single
photons produced by a QD are generally not
transform-limited, so that subsequent photon events are not
completely indistinguishable. The additional dephasing leads to a
photon coherence time that is $\approx$280 ps for these
QDs emitting at 1.3~$\mu$m \cite{ref:Srinivasan17}, yielding a photon indistinguishability of $\approx10.0$~$\%$ \cite{ref:Bylander_EPJD}.  In our
experiments, this timescale implies that while each photon is
modulated by the EOM, they are not all modulated in the same way as in Ref.~\onlinecite{ref:Kolchin_PRL_08} or \onlinecite{ref:Specht_NPhot_09}.  While this caveat is important to correctly interpret the data presented here, the methods and techniques would work just as well for a QD whose lifetime has been reduced by Purcell enhancement \cite{ref:Santori2,ref:Varoutsis_PRB05} in order to restore the indistinguishability completely.  Furthermore, electro-optic modulation could itself be used to improve the indistinguishability at the cost of losing some of the photons by post-selection \cite{ref:Patel_PRL08}, but without the narrow spectral bandwidth characteristic of the Purcell Effect.  In fact for the QD used in this work, with a lifetime of 1.4~ns and a photon coherence time of $\approx$280 ps, the indistinguishability could be increased to near-unity by using a Gaussian-shaped modulation with $\tau_{mod}\approx140$~ps.  The fraction of photons that would be lost due to the small temporal width would be $\approx$90~$\%$, yielding a maximum single photon count rate of $\approx6.8\times 10^4$ s$^{-1}$ after accounting for the collection efficiency of the FTW (see supporting material).  For a QD with a longer photon coherence time (580~ps was measured in Ref.~\onlinecite{ref:Flagg_PRL10}) the transmitted count rate can be much higher, resulting in a significant amount of identical single photons with sub-nanosecond pulse durations all in a single mode fiber.

In conclusion, we have clearly demonstrated sub-nanosecond modulation of triggered single photons from a quantum dot with a variety of waveforms.  Given the need for efficient interaction between single photons and local quantum systems, such single photon control will be a useful resource for quantum networks.  In addition, amplitude modulation may also prove to be useful for increasing the indistinguishability of triggered single photons from a quantum dot.

The authors acknowledge technical assistance from Alan Band at the CNST and useful discussions with S.~E. Harris at Stanford University.



\begin{thebibliography}{26}
\expandafter\ifx\csname natexlab\endcsname\relax\def\natexlab#1{#1}\fi
\expandafter\ifx\csname bibnamefont\endcsname\relax
  \def\bibnamefont#1{#1}\fi
\expandafter\ifx\csname bibfnamefont\endcsname\relax
  \def\bibfnamefont#1{#1}\fi
\expandafter\ifx\csname citenamefont\endcsname\relax
  \def\citenamefont#1{#1}\fi
\expandafter\ifx\csname url\endcsname\relax
  \def\url#1{\texttt{#1}}\fi
\expandafter\ifx\csname urlprefix\endcsname\relax\def\urlprefix{URL }\fi
\providecommand{\bibinfo}[2]{#2}
\providecommand{\eprint}[2][]{\url{#2}}

\bibitem[{\citenamefont{Gisin et~al.}(2002)\citenamefont{Gisin, Ribordy,
  Tittel, and Zbinden}}]{ref:Gisin_RMP_02}
\bibinfo{author}{\bibfnamefont{N.}~\bibnamefont{Gisin}},
  \bibinfo{author}{\bibfnamefont{G.}~\bibnamefont{Ribordy}},
  \bibinfo{author}{\bibfnamefont{W.}~\bibnamefont{Tittel}}, \bibnamefont{and}
  \bibinfo{author}{\bibfnamefont{H.}~\bibnamefont{Zbinden}},
  \bibinfo{journal}{Rev. Mod. Phys.} \textbf{\bibinfo{volume}{74}},
  \bibinfo{pages}{145} (\bibinfo{year}{2002}).

\bibitem[{\citenamefont{Knill et~al.}(2001)\citenamefont{Knill, Laflamme, and
  Milburn}}]{ref:Knill}
\bibinfo{author}{\bibfnamefont{E.}~\bibnamefont{Knill}},
  \bibinfo{author}{\bibfnamefont{R.}~\bibnamefont{Laflamme}}, \bibnamefont{and}
  \bibinfo{author}{\bibfnamefont{G.~J.} \bibnamefont{Milburn}},
  \bibinfo{journal}{Nature} \textbf{\bibinfo{volume}{409}}, \bibinfo{pages}{46}
  (\bibinfo{year}{2001}).

\bibitem[{\citenamefont{Raussendorf and Briegel}(2001)}]{ref:Raussendorf_PRL01}
\bibinfo{author}{\bibfnamefont{R.}~\bibnamefont{Raussendorf}} \bibnamefont{and}
  \bibinfo{author}{\bibfnamefont{H.~J.} \bibnamefont{Briegel}},
  \bibinfo{journal}{Phys. Rev. Lett.} \textbf{\bibinfo{volume}{86}},
  \bibinfo{pages}{5188} (\bibinfo{year}{2001}).

\bibitem[{\citenamefont{{Shields}}(2007)}]{ref:Shields_NPhot}
\bibinfo{author}{\bibfnamefont{A.~J.} \bibnamefont{{Shields}}},
  \bibinfo{journal}{Nature Photonics} \textbf{\bibinfo{volume}{1}},
  \bibinfo{pages}{215} (\bibinfo{year}{2007}).

\bibitem[{\citenamefont{Santori et~al.}(2001)\citenamefont{Santori, Pelton,
  Solomon, Dale, and Yamamoto}}]{ref:Santori}
\bibinfo{author}{\bibfnamefont{C.}~\bibnamefont{Santori}},
  \bibinfo{author}{\bibfnamefont{M.}~\bibnamefont{Pelton}},
  \bibinfo{author}{\bibfnamefont{G.}~\bibnamefont{Solomon}},
  \bibinfo{author}{\bibfnamefont{Y.}~\bibnamefont{Dale}}, \bibnamefont{and}
  \bibinfo{author}{\bibfnamefont{Y.}~\bibnamefont{Yamamoto}},
  \bibinfo{journal}{Phys. Rev. Lett.} \textbf{\bibinfo{volume}{86}},
  \bibinfo{pages}{1502} (\bibinfo{year}{2001}).

\bibitem[{\citenamefont{{Strauf} et~al.}(2007)\citenamefont{{Strauf}, {Stoltz},
  {Rakher}, {Coldren}, {Petroff}, and {Bouwmeester}}}]{ref:Strauf_NPhot}
\bibinfo{author}{\bibfnamefont{S.}~\bibnamefont{{Strauf}}},
  \bibinfo{author}{\bibfnamefont{N.~G.} \bibnamefont{{Stoltz}}},
  \bibinfo{author}{\bibfnamefont{M.~T.} \bibnamefont{{Rakher}}},
  \bibinfo{author}{\bibfnamefont{L.~A.} \bibnamefont{{Coldren}}},
  \bibinfo{author}{\bibfnamefont{P.~M.} \bibnamefont{{Petroff}}},
  \bibnamefont{and}
  \bibinfo{author}{\bibfnamefont{D.}~\bibnamefont{{Bouwmeester}}},
  \bibinfo{journal}{Nature Photonics} \textbf{\bibinfo{volume}{1}},
  \bibinfo{pages}{704} (\bibinfo{year}{2007}).

\bibitem[{\citenamefont{{Claudon} et~al.}(2010)\citenamefont{{Claudon},
  {Bleuse}, {Malik}, {Bazin}, {Jaffrennou}, {Gregersen}, {Sauvan}, {Lalanne},
  and {G{\'e}rard}}}]{ref:Claudon}
\bibinfo{author}{\bibfnamefont{J.}~\bibnamefont{{Claudon}}},
  \bibinfo{author}{\bibfnamefont{J.}~\bibnamefont{{Bleuse}}},
  \bibinfo{author}{\bibfnamefont{N.~S.} \bibnamefont{{Malik}}},
  \bibinfo{author}{\bibfnamefont{M.}~\bibnamefont{{Bazin}}},
  \bibinfo{author}{\bibfnamefont{P.}~\bibnamefont{{Jaffrennou}}},
  \bibinfo{author}{\bibfnamefont{N.}~\bibnamefont{{Gregersen}}},
  \bibinfo{author}{\bibfnamefont{C.}~\bibnamefont{{Sauvan}}},
  \bibinfo{author}{\bibfnamefont{P.}~\bibnamefont{{Lalanne}}},
  \bibnamefont{and}
  \bibinfo{author}{\bibfnamefont{J.}~\bibnamefont{{G{\'e}rard}}},
  \bibinfo{journal}{Nature Photonics} \textbf{\bibinfo{volume}{4}},
  \bibinfo{pages}{174} (\bibinfo{year}{2010}).

\bibitem[{\citenamefont{Davan\c{c}o and Srinivasan}(2009)}]{ref:Davanco2}
\bibinfo{author}{\bibfnamefont{M.}~\bibnamefont{Davan\c{c}o}} \bibnamefont{and}
  \bibinfo{author}{\bibfnamefont{K.}~\bibnamefont{Srinivasan}},
  \bibinfo{journal}{Opt. Lett.} \textbf{\bibinfo{volume}{34}},
  \bibinfo{pages}{2542} (\bibinfo{year}{2009}).

\bibitem[{\citenamefont{Cirac et~al.}(1997)\citenamefont{Cirac, Zoller, Kimble,
  and Mabuchi}}]{ref:Cirac}
\bibinfo{author}{\bibfnamefont{J.~I.} \bibnamefont{Cirac}},
  \bibinfo{author}{\bibfnamefont{P.}~\bibnamefont{Zoller}},
  \bibinfo{author}{\bibfnamefont{H.~J.} \bibnamefont{Kimble}},
  \bibnamefont{and} \bibinfo{author}{\bibfnamefont{H.}~\bibnamefont{Mabuchi}},
  \bibinfo{journal}{Phys. Rev. Lett.} \textbf{\bibinfo{volume}{78}},
  \bibinfo{pages}{3221} (\bibinfo{year}{1997}).

\bibitem[{\citenamefont{Gorshkov et~al.}(2007)\citenamefont{Gorshkov, Andr\'e,
  Fleischhauer, S\o{}rensen, and Lukin}}]{ref:Gorshkov_PRL07}
\bibinfo{author}{\bibfnamefont{A.~V.} \bibnamefont{Gorshkov}},
  \bibinfo{author}{\bibfnamefont{A.}~\bibnamefont{Andr\'e}},
  \bibinfo{author}{\bibfnamefont{M.}~\bibnamefont{Fleischhauer}},
  \bibinfo{author}{\bibfnamefont{A.~S.} \bibnamefont{S\o{}rensen}},
  \bibnamefont{and} \bibinfo{author}{\bibfnamefont{M.~D.} \bibnamefont{Lukin}},
  \bibinfo{journal}{Phys. Rev. Lett.} \textbf{\bibinfo{volume}{98}},
  \bibinfo{pages}{123601} (\bibinfo{year}{2007}).

\bibitem[{\citenamefont{{Kimble}}(2008)}]{ref:Kimble_Nat08}
\bibinfo{author}{\bibfnamefont{H.~J.} \bibnamefont{{Kimble}}},
  \bibinfo{journal}{\nat} \textbf{\bibinfo{volume}{453}}, \bibinfo{pages}{1023}
  (\bibinfo{year}{2008}).

\bibitem[{\citenamefont{{Kolchin} et~al.}(2008)\citenamefont{{Kolchin},
  {Belthangady}, {Du}, {Yin}, and {Harris}}}]{ref:Kolchin_PRL_08}
\bibinfo{author}{\bibfnamefont{P.}~\bibnamefont{{Kolchin}}},
  \bibinfo{author}{\bibfnamefont{C.}~\bibnamefont{{Belthangady}}},
  \bibinfo{author}{\bibfnamefont{S.}~\bibnamefont{{Du}}},
  \bibinfo{author}{\bibfnamefont{G.~Y.} \bibnamefont{{Yin}}}, \bibnamefont{and}
  \bibinfo{author}{\bibfnamefont{S.~E.} \bibnamefont{{Harris}}},
  \bibinfo{journal}{Phys. Rev. Lett.} \textbf{\bibinfo{volume}{101}},
  \bibinfo{pages}{103601} (\bibinfo{year}{2008}).

\bibitem[{\citenamefont{{Specht} et~al.}(2009)\citenamefont{{Specht},
  {Bochmann}, {M{\"u}cke}, {Weber}, {Figueroa}, {Moehring}, and
  {Rempe}}}]{ref:Specht_NPhot_09}
\bibinfo{author}{\bibfnamefont{H.~P.} \bibnamefont{{Specht}}},
  \bibinfo{author}{\bibfnamefont{J.}~\bibnamefont{{Bochmann}}},
  \bibinfo{author}{\bibfnamefont{M.}~\bibnamefont{{M{\"u}cke}}},
  \bibinfo{author}{\bibfnamefont{B.}~\bibnamefont{{Weber}}},
  \bibinfo{author}{\bibfnamefont{E.}~\bibnamefont{{Figueroa}}},
  \bibinfo{author}{\bibfnamefont{D.~L.} \bibnamefont{{Moehring}}},
  \bibnamefont{and} \bibinfo{author}{\bibfnamefont{G.}~\bibnamefont{{Rempe}}},
  \bibinfo{journal}{Nature Photonics} \textbf{\bibinfo{volume}{3}},
  \bibinfo{pages}{469} (\bibinfo{year}{2009}).

\bibitem[{\citenamefont{McKeever et~al.}(2004)\citenamefont{McKeever, Boca,
  Boozer, Miller, Buck, Kuzmich, and Kimble}}]{ref:McKeever}
\bibinfo{author}{\bibfnamefont{J.}~\bibnamefont{McKeever}},
  \bibinfo{author}{\bibfnamefont{A.}~\bibnamefont{Boca}},
  \bibinfo{author}{\bibfnamefont{A.~D.} \bibnamefont{Boozer}},
  \bibinfo{author}{\bibfnamefont{R.}~\bibnamefont{Miller}},
  \bibinfo{author}{\bibfnamefont{J.~R.} \bibnamefont{Buck}},
  \bibinfo{author}{\bibfnamefont{A.}~\bibnamefont{Kuzmich}}, \bibnamefont{and}
  \bibinfo{author}{\bibfnamefont{H.~J.} \bibnamefont{Kimble}},
  \bibinfo{journal}{Science} \textbf{\bibinfo{volume}{303}},
  \bibinfo{pages}{1992} (\bibinfo{year}{2004}).

\bibitem[{\citenamefont{Keller et~al.}(2004)\citenamefont{Keller, Lange,
  Hayaska, Lange, and Walther}}]{ref:Keller}
\bibinfo{author}{\bibfnamefont{M.}~\bibnamefont{Keller}},
  \bibinfo{author}{\bibfnamefont{B.}~\bibnamefont{Lange}},
  \bibinfo{author}{\bibfnamefont{K.}~\bibnamefont{Hayaska}},
  \bibinfo{author}{\bibfnamefont{W.}~\bibnamefont{Lange}}, \bibnamefont{and}
  \bibinfo{author}{\bibfnamefont{H.}~\bibnamefont{Walther}},
  \bibinfo{journal}{Nature} \textbf{\bibinfo{volume}{431}},
  \bibinfo{pages}{1075} (\bibinfo{year}{2004}).

\bibitem[{\citenamefont{Srinivasan et~al.}(2007)\citenamefont{Srinivasan,
  Painter, Stintz, and Krishna}}]{ref:Srinivasan15}
\bibinfo{author}{\bibfnamefont{K.}~\bibnamefont{Srinivasan}},
  \bibinfo{author}{\bibfnamefont{O.}~\bibnamefont{Painter}},
  \bibinfo{author}{\bibfnamefont{A.}~\bibnamefont{Stintz}}, \bibnamefont{and}
  \bibinfo{author}{\bibfnamefont{S.}~\bibnamefont{Krishna}},
  \bibinfo{journal}{Appl. Phys. Lett.} \textbf{\bibinfo{volume}{91}},
  \bibinfo{pages}{091102} (\bibinfo{year}{2007}).

\bibitem[{\citenamefont{{Rakher} et~al.}(2010)\citenamefont{{Rakher}, {Ma},
  {Slattery}, {Tang}, and {Srinivasan}}}]{ref:Rakher_NPhot_2010}
\bibinfo{author}{\bibfnamefont{M.~T.} \bibnamefont{{Rakher}}},
  \bibinfo{author}{\bibfnamefont{L.}~\bibnamefont{{Ma}}},
  \bibinfo{author}{\bibfnamefont{O.}~\bibnamefont{{Slattery}}},
  \bibinfo{author}{\bibfnamefont{X.}~\bibnamefont{{Tang}}}, \bibnamefont{and}
  \bibinfo{author}{\bibfnamefont{K.}~\bibnamefont{{Srinivasan}}},
  \bibinfo{journal}{Nature Photonics} \textbf{\bibinfo{volume}{4}},
  \bibinfo{pages}{786} (\bibinfo{year}{2010}).

\bibitem[{\citenamefont{{Hadfield}}(2009)}]{ref:Hadfield_nphoton_09}
\bibinfo{author}{\bibfnamefont{R.~H.} \bibnamefont{{Hadfield}}},
  \bibinfo{journal}{Nature Photonics} \textbf{\bibinfo{volume}{3}},
  \bibinfo{pages}{696} (\bibinfo{year}{2009}).

\bibitem[{\citenamefont{Broadbent et~al.}(2009)\citenamefont{Broadbent, Zerom,
  Shin, Howell, and Boyd}}]{ref:Broadbent_PRA09}
\bibinfo{author}{\bibfnamefont{C.~J.} \bibnamefont{Broadbent}},
  \bibinfo{author}{\bibfnamefont{P.}~\bibnamefont{Zerom}},
  \bibinfo{author}{\bibfnamefont{H.}~\bibnamefont{Shin}},
  \bibinfo{author}{\bibfnamefont{J.~C.} \bibnamefont{Howell}},
  \bibnamefont{and} \bibinfo{author}{\bibfnamefont{R.~W.} \bibnamefont{Boyd}},
  \bibinfo{journal}{Phys. Rev. A} \textbf{\bibinfo{volume}{79}},
  \bibinfo{pages}{033802} (\bibinfo{year}{2009}).

\bibitem[{\citenamefont{Belthangady et~al.}(2010)\citenamefont{Belthangady,
  Chuu, Yu, Yin, Kahn, and Harris}}]{ref:Belthangady_PRL10}
\bibinfo{author}{\bibfnamefont{C.}~\bibnamefont{Belthangady}},
  \bibinfo{author}{\bibfnamefont{C.-S.} \bibnamefont{Chuu}},
  \bibinfo{author}{\bibfnamefont{I.~A.} \bibnamefont{Yu}},
  \bibinfo{author}{\bibfnamefont{G.~Y.} \bibnamefont{Yin}},
  \bibinfo{author}{\bibfnamefont{J.~M.} \bibnamefont{Kahn}}, \bibnamefont{and}
  \bibinfo{author}{\bibfnamefont{S.~E.} \bibnamefont{Harris}},
  \bibinfo{journal}{Phys. Rev. Lett.} \textbf{\bibinfo{volume}{104}},
  \bibinfo{pages}{223601} (\bibinfo{year}{2010}).

\bibitem[{\citenamefont{Srinivasan et~al.}(2008)\citenamefont{Srinivasan,
  Michael, Perahia, and Painter}}]{ref:Srinivasan17}
\bibinfo{author}{\bibfnamefont{K.}~\bibnamefont{Srinivasan}},
  \bibinfo{author}{\bibfnamefont{C.~P.} \bibnamefont{Michael}},
  \bibinfo{author}{\bibfnamefont{R.}~\bibnamefont{Perahia}}, \bibnamefont{and}
  \bibinfo{author}{\bibfnamefont{O.}~\bibnamefont{Painter}},
  \bibinfo{journal}{Phys. Rev. A} \textbf{\bibinfo{volume}{78}},
  \bibinfo{pages}{033839} (\bibinfo{year}{2008}).

\bibitem[{\citenamefont{{Bylander} et~al.}(2003)\citenamefont{{Bylander},
  {Robert-Philip}, and {Abram}}}]{ref:Bylander_EPJD}
\bibinfo{author}{\bibfnamefont{J.}~\bibnamefont{{Bylander}}},
  \bibinfo{author}{\bibfnamefont{I.}~\bibnamefont{{Robert-Philip}}},
  \bibnamefont{and} \bibinfo{author}{\bibfnamefont{I.}~\bibnamefont{{Abram}}},
  \bibinfo{journal}{Eur. Phys. J. D} \textbf{\bibinfo{volume}{22}},
  \bibinfo{pages}{295} (\bibinfo{year}{2003}).

\bibitem[{\citenamefont{Santori et~al.}(2002)\citenamefont{Santori, Fattal,
  Vuckovic, Solomon, and Yamamoto}}]{ref:Santori2}
\bibinfo{author}{\bibfnamefont{C.}~\bibnamefont{Santori}},
  \bibinfo{author}{\bibfnamefont{D.}~\bibnamefont{Fattal}},
  \bibinfo{author}{\bibfnamefont{J.}~\bibnamefont{Vuckovic}},
  \bibinfo{author}{\bibfnamefont{G.}~\bibnamefont{Solomon}}, \bibnamefont{and}
  \bibinfo{author}{\bibfnamefont{Y.}~\bibnamefont{Yamamoto}},
  \bibinfo{journal}{Nature} \textbf{\bibinfo{volume}{419}},
  \bibinfo{pages}{594} (\bibinfo{year}{2002}).

\bibitem[{\citenamefont{Varoutsis et~al.}(2005)\citenamefont{Varoutsis,
  Laurent, Kramper, Lema\^\i{}tre, Sagnes, Robert-Philip, and
  Abram}}]{ref:Varoutsis_PRB05}
\bibinfo{author}{\bibfnamefont{S.}~\bibnamefont{Varoutsis}},
  \bibinfo{author}{\bibfnamefont{S.}~\bibnamefont{Laurent}},
  \bibinfo{author}{\bibfnamefont{P.}~\bibnamefont{Kramper}},
  \bibinfo{author}{\bibfnamefont{A.}~\bibnamefont{Lema\^\i{}tre}},
  \bibinfo{author}{\bibfnamefont{I.}~\bibnamefont{Sagnes}},
  \bibinfo{author}{\bibfnamefont{I.}~\bibnamefont{Robert-Philip}},
  \bibnamefont{and} \bibinfo{author}{\bibfnamefont{I.}~\bibnamefont{Abram}},
  \bibinfo{journal}{Phys. Rev. B} \textbf{\bibinfo{volume}{72}},
  \bibinfo{pages}{041303} (\bibinfo{year}{2005}).

\bibitem[{\citenamefont{Patel et~al.}(2008)\citenamefont{Patel, Bennett,
  Cooper, Atkinson, Nicoll, Ritchie, and Shields}}]{ref:Patel_PRL08}
\bibinfo{author}{\bibfnamefont{R.~B.} \bibnamefont{Patel}},
  \bibinfo{author}{\bibfnamefont{A.~J.} \bibnamefont{Bennett}},
  \bibinfo{author}{\bibfnamefont{K.}~\bibnamefont{Cooper}},
  \bibinfo{author}{\bibfnamefont{P.}~\bibnamefont{Atkinson}},
  \bibinfo{author}{\bibfnamefont{C.~A.} \bibnamefont{Nicoll}},
  \bibinfo{author}{\bibfnamefont{D.~A.} \bibnamefont{Ritchie}},
  \bibnamefont{and} \bibinfo{author}{\bibfnamefont{A.~J.}
  \bibnamefont{Shields}}, \bibinfo{journal}{Phys. Rev. Lett.}
  \textbf{\bibinfo{volume}{100}}, \bibinfo{pages}{207405}
  (\bibinfo{year}{2008}).

\bibitem[{\citenamefont{Flagg et~al.}(2010)\citenamefont{Flagg, Muller,
  Polyakov, Ling, Migdall, and Solomon}}]{ref:Flagg_PRL10}
\bibinfo{author}{\bibfnamefont{E.~B.} \bibnamefont{Flagg}},
  \bibinfo{author}{\bibfnamefont{A.}~\bibnamefont{Muller}},
  \bibinfo{author}{\bibfnamefont{S.~V.} \bibnamefont{Polyakov}},
  \bibinfo{author}{\bibfnamefont{A.}~\bibnamefont{Ling}},
  \bibinfo{author}{\bibfnamefont{A.}~\bibnamefont{Migdall}}, \bibnamefont{and}
  \bibinfo{author}{\bibfnamefont{G.~S.} \bibnamefont{Solomon}},
  \bibinfo{journal}{Phys. Rev. Lett.} \textbf{\bibinfo{volume}{104}},
  \bibinfo{pages}{137401} (\bibinfo{year}{2010}).

\end{thebibliography}
\end{document}